\documentclass{svjour2}                    
\smartqed  
\usepackage{graphicx}
\newcommand{\trace}[1]{\ensuremath{\mathrm{Tr}\left[{#1}\right]}}

\newcommand{\partrace}[2]{\ensuremath{\mathrm{Tr}_{{#2}}\left[{#1}\right]}}

\newcommand{\func}[1]{\ensuremath{\left[ {#1} \right]}}

\begin{document}
\title{Does a Computer have an Arrow of Time?}

\author{Owen J E Maroney}

\institute{
Centre for Time, SOPHI, University of Sydney, NSW 2006, Australia\\
\and
Perimeter Institute for Theoretical Physics, 31 Caroline Street North, Waterloo,
Ontario, N2L 2Y5, Canada \\
\email{o.maroney@usyd.edu.au}}
\date{Received: date / Accepted: date}

\maketitle

\begin{abstract}Schulman\cite{Schulman2005a} has argued that Boltzmann's intuition,
that the psychological arrow of time is necessarily aligned with the
thermodynamic arrow, is correct.  Schulman gives an explicit
physical mechanism for this connection, based on the brain being
representable as a computer, together with certain thermodynamic
properties of computational processes. Hawking\cite{Hawking1994} presents
similar, if briefer, arguments.  The purpose of this paper is to critically examine the support for
the link between thermodynamics and an arrow of time for computers.
The principal arguments put forward by Schulman and Hawking will be
shown to fail. It will be shown that any computational process that
can take place in an entropy increasing universe, can equally take
place in an entropy decreasing universe.  This conclusion does not automatically imply a psychological arrow
can run counter to the thermodynamic arrow.  Some alternative
possible explanations for the alignment of the two arrows will be
briefly discussed.
\end{abstract}

\section{Introduction}
In part of his response to Zermelo's reversibility objections to
statistical mechanics, Boltzmann\cite{Boltzmann1898} suggested it
was possible (indeed, inevitable) to have extended regions of space,
and time, that were entropy decreasing, but that living beings
within those regions would be unable to perceive the difference:
\begin{quote}
For the universe, the two directions of time are indistinguishable,
just as in space there is no up and down.  However, just as at a
particular place on the earth's surface we call ``down'' the
direction toward the center of the earth, so will a living being in
a particular time interval of such a single world distinguish the
direction of time toward the less probable state from the opposite
direction (the former toward the past, the latter toward the future)
\end{quote}
Authors such as\cite{Reichenbach1971,Horwich1987} have developed
this idea while others\cite{Skl93,Earman2006,Maudlin2002} are
critical.

As noted in \cite{Sklar1985}[Chapter 12], the perception of `up' and
`down' can be directly traced to particular physical processes in
different creatures (and specifically in the case of humans, the
effect of the gravitational field on the fluid of the inner ear).
While it may seem implausible that there could be an equivalent
organ, which monitors the local entropy gradient, and informs the
brain in which direction time is flowing, there remains the
possibility that there is still something about the general
functioning of the brain that can only take place in the direction
of entropy increase.

In a recent paper Schulman\cite{Schulman2005a} claims to identify
such a function from the general thermodynamic properties of
computations, as physical processes.  He gives a detailed comparison
of the components of a computer with the features of the
psychological arrow to show
\begin{quote}
the extent to which a computer \ldots can be said to possess a
psychological arrow.  My contention is that the parallels are
sufficiently strong as to leave little room for an independent
psychological arrow.
\end{quote}
He then appeals to Landauer's Principle\cite{Lan61} to show that the
intrinsic arrow of computational processes must be aligned with the
thermodynamic arrow.  As a result a computer is
\begin{quote}
without an independent arrow of time, retaining the past/future
distinction by virtue of its being part of a mechanistic world with
a thermodynamic arrow in a particular direction.
\end{quote}
Similar suggestions to Schulman's can be found in \cite{Hawking1985,Hawking1987,Hawking1994}
\begin{quote}
when a computer records something in memory, the total entropy
increases.  Thus computers remember things in the direction of time
in which entropy increases.  In a universe in which entropy is
decreasing in time, computer memories will work backward.\cite{Hawking1994}
\end{quote}
It is argued in this paper that neither Hawking nor Schulman's
arguments hold.

The structure is as follows. First (Section
\ref{s:thdynarrow}) we will state how we will treat the
thermodynamic arrow of time, and what we mean when we refer to an
`entropy increasing universe' and an `entropy decreasing universe'.
Then (Section \ref{s:trevcomp}) we consider what it takes for a
physical process to embody a computation and the effect of a time
reversal of this physical process.  The processes that result from
this temporal reversal are not equivalent to the processes that can
represent a computation. We then show the key result that equivalent
operations to the time reversed processes can be constructed, so the
time reverse of those equivalent operations is a computation in a
time reversed universe (Section \ref{ss:trevseq}) that is equivalent
to the original computation. This demonstrates the physical
possibility of such processes in entropy decreasing universes, and
gives us a model to further study the possibilities of computation
under such circumstances.

In Section \ref{s:landp}, we examine the derivation of Landauer's
Principle in an entropy decreasing universe.  We find that the
physical assumptions required for an entropy decreasing universe
result in a reversal of the inequality that occurs in the usual
statements of Landauer's Principle.  Rather than necessitating
entropy increases, when taking place in an entropy decreasing
universe logical operations necessitate entropy decreases.  In
retrospect this will seem rather obvious.

Finally (Section \ref{s:corr}) we consider the question of whether
systems which gather, process and utilise information, are simply
more likely to arise in entropy decreasing or entropy increasing
universes.  We examine this from the point of view of volume of
state space arguments, to see if there is, all else being equal, any
reason to expect that entropy decreasing universes are inherently
hostile to the gathering and retention of information. We find that,
perhaps surprisingly, they are not.  We conclude that, on the basis
of statistical mechanical arguments alone, we have no grounds for
linking any computational arrow of time to the thermodynamic arrow
of time.

Given the clear manner in which our own information processing seems
aligned to the thermodynamic arrow, this may seem puzzling.  We will
briefly consider some possible explanations of this link, but which
would require more complex arguments to justify.    A surprising
conclusion might be that, if the psychological arrow of time is
necessarily aligned with the thermodynamic arrow, then it cannot be
logically supervenient upon computational states.  Alternatively, if
the psychological arrow of time is logically supervenient upon
information processing, then it must be logically independent of the
thermodynamic arrow.

\section{The Thermodynamic Arrow}\label{s:thdynarrow}
First it is necessary to make clear what is meant by an entropy
increasing universe and an entropy decreasing universe.

 The state space
of the universe is formed from the product of the state spaces of a
large number of smaller systems $\Omega = \prod_i \otimes \Omega_i$
and a measure, $\mu$, on regions of the state space. It will be
usually only be necessary to consider grouping the subsystems into a
small number of distinct, larger subsystems, $j$, with most of the
small subsystems grouped into a single `environment', $E$:
\begin{eqnarray}
\Omega_j &=& \prod_{i \in j} \otimes \Omega_i\\
\Omega_E &=& \prod_{i \in E} \otimes \Omega_i\\
\Omega &=& \Omega_E \prod_j \otimes \Omega_j
\end{eqnarray}

The dynamics are described by an invertible, measure preserving flow
$\phi^{(t)}$ on the state space.  For any region $\Delta \subseteq
\Omega$ then $\mu(\phi^{(t)}(\Delta))=\mu(\Delta)$, and there exists
a map $\phi^{-(t)}$ such that $\phi^{-(t)} \circ
\phi^{(t)}(\Delta)=\phi^{(t)} \circ \phi^{-(t)}(\Delta)=\Delta$.

\subsection{Entropy increasing universe}\label{ss:entinc}
An entropy increasing universe has a microstate that starts in a
very small and special region $\Delta_0 \subseteq \Omega$. It is
assumed that the dynamics of the flow on the state space is such
that, over time, this region spreads out over the state space. As
the measure is preserved, this can only happen by the region
developing a very elongated and filamentary structure.  As part of
the special nature of the initial region, it will be assumed that
the fine detail of this elongated and filamentary structure can be
ignored for any future evolution of the system.

The initial region is a direct product of regions over the
subsystems:
\[
\Delta_0 = \prod_i \otimes \Delta_i
\]
After the system has evolved, it will not, in general be the case
that the evolved region $\phi^{(t)}(\Delta_0)$ is a direct product
of regions over the subsystems.  Instead there will be microscopic correlations.  The appearance of entropy increase will be represented by the coarse graining out of these microscopic correlations.

We will assume that the state space $\Omega_i$ of each subsystem,
$i$, is divided into distinct subregions $\omega_{i,j}$, such that
$\cup_j \omega_{i,j}=\Omega_i$.  A direct product of a set of subregions across all the subsystems:
\begin{equation}
\omega_{\underline{n}}=\prod_i \otimes \omega_{i,n_i}
\end{equation}
can be represented by the array of integers $\underline{n}=\left(n_1, \ldots, n_j, \ldots \right)$ .

$\overline{\phi^{(t)}(\Delta_0)}$ is the coarse grained representation of $\phi^{(t)}(\Delta_0)$.  This is defined to be the smallest superset of $\phi^{(t)}(\Delta_0)$ that can be represented by a union of some set, $\{\underline{n}\}$, of direct products over the subsystems and a direct product with a subset of the environment.  It will be of the form
\begin{equation}
\phi^{(t)}(\Delta_0) \subseteq \overline{\phi^{(t)}(\Delta_0)}=\cup_{\underline{n}} \overline{\Delta}_{\underline{n}} \otimes \overline{\Delta}_E
\end{equation}

Let the sets $\left\{\Delta^\prime_{i,m_i}\right\}$ and
$\left\{\Delta^\prime_E\right\}$ be the sets of all the regions
that satisfy:
\begin{eqnarray}
\Delta^\prime_{i,m_i} &\subseteq& \omega_{i,m_i}\\
\Delta^\prime_E &\subseteq &\Omega_E
\end{eqnarray}
and for which there exits some set $\{\underline{m}\}$ such that
\begin{eqnarray}
\Delta^\prime_{\underline{m}} &=& \prod_i \otimes \Delta^\prime_{i,m_i}\\
\phi^{(t)}(\Delta_0) &\subseteq& \cup_{\underline{m}} \Delta^\prime_{\underline{m}} \otimes \Delta^\prime_E
\end{eqnarray}

The set $\{\underline{n}\}$ and subregions $\overline{\Delta}_{i,n_i} \in
\left\{\Delta^\prime_{i,m_i}\right\}$ and $\overline{\Delta}_E \in
\left\{\Delta^\prime_E\right\}$
that satisfy the conditions
\begin{eqnarray}
\overline{\Delta}_{\underline{n}} &=& \prod_i \otimes \overline{\Delta}_{i,n_i}\\
\phi^{(t)}(\Delta_0) &\subseteq& \cup_{\underline{n}} \overline{\Delta}_{\underline{n}} \otimes \overline{\Delta}_E \\
\cup_{\underline{n}} \overline{\Delta}_{\underline{n}} \otimes \overline{\Delta}_E
&\subseteq &\cup_{\underline{m}} \Delta^\prime_{\underline{m}}
\otimes \Delta^\prime_E
\end{eqnarray}
$\forall
$\{\underline{m}\}$, \Delta^\prime_E$, give the coarse graining of $\phi^{(t)}(\Delta_0)$.

In an entropy increasing universe, we assume that the microscopic
correlations that develop due to $\phi^{(t)}$ play no role in the
future evolution of the system.  In effect, this means that we may
make the coarse grained replacement
\begin{equation}
\phi^{(t)}(\Delta_0) \rightarrow \cup_{\underline{n}}
\overline{\Delta}_{\underline{n}} \otimes \overline{\Delta}_E
\end{equation}
for all future evolution of the system.

The requirement that the initial state $\Delta_0$ is such that it
produces all these results, for all realistic maps $\phi^{(t)}$,
will be referred to as the initial boundary condition, and the
resulting evolution as being in an entropy increasing universe. For
the purposes of this paper it will be assumed that these conditions
can be met.

When looking at the interactions of localised systems at times long
after the initial boundary condition, but long before complete
thermalisation (which occurs at some future time $t_{th}$), this is
represented by:
\begin{enumerate}
\item No initial microscopic correlations between macroscopic
subsystems;
\item Thermal states are represented by Gibbs distributions at the
start of any interaction.
\item Microscopic correlations develop between the subsystems;
\item The sum of the Gibbs entropies of the marginal distributions of the
macroscopic subsystems, increases;
\item The microscopic correlations become, for all practical
purposes, inaccessible and may be coarse grained away;
\end{enumerate}

\subsection{Time reversal and symmetry}
For clarity, we now state explicitly what we will mean by time
reversal and by time symmetries.

The time reversal of a dynamical system involves taking the time ordered sequence of regions, $\Delta_t$, generated by the flow $\Delta_t=\phi^{(t)}(\Delta_0)$, time reversing each individual state, then reversing the order in which the states occur.  This generates a new time ordered sequence of regions $\Delta_{Tt}$.

First we need the notion of
the time reversal of the state space.  This is not unproblematic
(see \cite{Alb01}[Chapter 1], for example) but for the purposes of
this article let us assume that there is no disagreement over the
time reverse of a state in our state space.  The time reversal of
the state space is a map $\Delta^T=T(\Delta) \subset \Omega$ such
that $\mu(\Delta^T)=\mu(\Delta)$ and $\Delta=T \circ T(\Delta)$. For
subsystems $T(\prod_i \otimes \Delta_i)=\prod_i \otimes T(\Delta_i)$
and for subspaces $T(\cup_n \Delta_n)=\cup_n T(\Delta_n)$.  We also
note if $A \subset B$ then $A^T \subset B^T$ and for all state
spaces $\Omega$ we consider here $\Omega^T=\Omega$.

Reversing the order in which the states occur, about the time $t=t_0$,
corresponds\footnote{We need: $t=t_0 \rightarrow t=t_0$ and $t=0 \rightarrow t=2t_0$.} to $t \rightarrow 2 t_0-t$.  Combined we have the sequence of regions $\Delta_{Tt}=T(\Delta_{2t_0-t})=T \circ \phi^{(2t_0-t)}(\Delta_0)$.  At $t=0$ the sequence is at $\Delta_{T0}=T \circ \phi^{(2t_0)}(\Delta_0)$ so $\Delta_0=\phi^{-(2t_0)} \circ T(\Delta_{T0})$.  Combined, we have the time reversed dynamical map $\phi_{Tt_0}^{(t)}(\Delta)$:
\begin{equation}
\phi_{Tt_0}^{(t)}(\Delta)=T \circ \phi^{(2t_0-t)} \circ
\phi^{-(2t_0)} \circ T(\Delta)
\end{equation}
Two special cases may be more familiar.  Firstly, for $t_0=0$ we
have
\[
\phi_{T0}^{(t)}(\Delta)=T \circ \phi^{(-t)} \circ T(\Delta)
\]
Secondly, for a transformation $\phi^{(2t_0)}$, which takes place
over the time period $0<t<2t_0$, then a reversal at $t=t_0$ has the
transformation
\[
\phi_{Tt_0}^{(2t_0)}(\Delta)=T \circ \phi^{-(2t_0)} \circ T(\Delta)
\]

It is important to note one cannot use the coarse grained
description $\cup_{\underline{x}} \Delta_{\underline{x}} \otimes
\Delta_E$, defined in the previous section, for the time reversed
dynamics.  This coarse graining is valid, in the original dynamics,
only for later times so is valid only for earlier times in the time
reversed dynamics.

We now define time reversal invariance and time translation
invariance of the dynamics, although unless explicitly stated, we
will not be assuming any of these invariances hold.  We explicitly
state them so that it may be clear where we have not needed to
assume them.

The dynamics are time reversal invariant at $t_0$ iff
\begin{equation}\phi_{Tt_0}^{(t)}(\Delta)=\phi^{(t)}(\Delta)\end{equation}

Weak time translation invariance is defined as
\begin{equation}
\forall t>0,s > 0 \;\;\; \phi^{(t)} \circ
\phi^{(s)}(\Delta)=\phi^{(t+s)}(\Delta)
\end{equation}
and strong time translation invariance as
\begin{equation}
\forall t,s  \;\;\; \phi^{(t)} \circ
\phi^{(s)}(\Delta)=\phi^{(t+s)}(\Delta)
\end{equation}
Strong time translation invariance implies\footnote{As stated
previously, $\phi^{(t)}$ is invertible.} $\phi^{-(t)} =\phi^{(-t)}$,
and this in turn implies $\phi_{Tt_0}^{(t)}(\Delta)=T \circ
\phi^{(-t)} \circ T (\Delta)$ for all $t_0$.

If a dynamics is time reversal invariant at all times, it is
necessarily strong time translation invariant:
\begin{equation}
\left( \forall t_0 \;\;\;
\phi_{Tt_0}^{(t)}(\Delta)=\phi^{(t)}(\Delta) \right) \Rightarrow
\left( \forall t,s \;\;\; \phi^{(t)} \circ
\phi^{(s)}(\Delta)=\phi^{(t+s)}(\Delta) \right)
\end{equation}

If a dynamics is strong time translation invariant and time reversal
invariant at a single time, then it is necessarily time reversal
invariant at all times.
\begin{equation}
\left( \left( \forall t,s \;\;\; \phi^{(t)} \circ
\phi^{(s)}(\Delta)=\phi^{(t+s)}(\Delta) \right) \& \left( \exists
t_0 | \phi_{Tt_0}^{(t)}(\Delta)=\phi^{(t)}(\Delta) \right) \right)
\Rightarrow \left( \forall t_0 \;\;\;
\phi_{Tt_0}^{(t)}(\Delta)=\phi^{(t)}(\Delta) \right)
\end{equation}

\subsection{Entropy decreasing universe}\label{ss:entdec}
Now we must consider what it means to be in an entropy decreasing universe.  Broadly, this must exhibit the time reversal of the behaviour that defines an entropy increasing universe.  We must postulate the existence of a
future boundary condition, that at some future time $\tau$, the
state of the universe will be in the region of state space
$\Delta_0^T$ and with a suitable dynamics.

The most general means of doing this is to consider the
time reversal, at $\tau/2$, of entropy increasing universes.  Take the time ordered sequence of regions of state space in entropy increasing universes, $\Delta_t=\phi^{(t)}(\Delta_0)$, with final region $\Delta_\tau=\phi^{(\tau)}(\Delta_0)$.
The time reversed dynamics is
\begin{equation}
\Delta_{Tt}=\phi_{T\tau/2}^{(t)}(\Delta_{T0})=T\circ \phi^{(\tau-t)}\circ \phi^{-(\tau)} \circ T (\Delta_{T0})
\end{equation}
The initial condition, in the entropy decreasing universes, is the time reversal of the final condition of the entropy increasing universes: $\Delta_{T0}=T(\Delta_\tau)=T \circ \phi^{(\tau)}(\Delta_0)$.  So the time ordered sequence of regions of entropy decreasing universes is given by:
\begin{equation}
\Delta_{Tt}=T\circ \phi^{(\tau-t)}(\Delta_{0})
\end{equation}
If the dynamics are time reversal
invariant at $\tau/2$, then
$\phi_{T\tau/2}^{(t)}=\phi^{(t)}$ and $\Delta_{Tt}=\phi^{(t)}\circ T\circ \phi^{(\tau)}(\Delta_{0})$.

The coarse graining now works in reverse.  Over the
course of the evolution of the system, fine grained structure, of an
elongated and filamentary kind, appears.  This fine grained
structure played no role in the evolution of the system prior to its
appearance.  However, its appearance allows the region of state
space to evolve into smaller regions that its initial, coarse
grained, appearance would have indicated.  In thermodynamic terms,
this can be characterised by a universal tendency for heat to
spontaneously flow out of the environment and cause masses to be
raised through gravitational potentials.

When looking at the interactions of localised systems at times long
before the future boundary condition, $t=\tau$, but long after the
universe has come out of complete thermalisation, $t=\tau-t_{th}$,
this will be represented by the reversed set of conditions:
\begin{enumerate}
\item A high degree of initial microscopic correlations between macroscopic
subsystems.
\item Microscopic correlations disappear over the course of the interaction;
\item The sum of the Gibbs entropies of the marginal distributions of the
macroscopic subsystems, is decreasing;
\item The microscopic correlations which disappear, played no role in the earlier evolution of the
system.  In the future evolution of the system, new microscopic
correlations come into play;
\item Thermal states are represented by Gibbs distributions at the
{\it end} of any interaction.
\end{enumerate}

We have been considering regions of state space, and the time ordered sequences of regions of state space.  Any given universe follows only a single trajectory, a time ordered sequence of individual states.  Reversing the trajectory of a given universe necessarily reverses every single time asymmetric arrow associated with that universe.  When asking whether a particular time asymmetric arrow is a consequence of the thermodynamic arrow, we cannot answer the question by simply time reversing the trajectory of our particular universe.  Instead we must look at all the trajectories associated with the time ordered sequences of regions in the entropy increasing and decreasing cases, and then see if the arrow we are interested in turns out to have the same alignment to the thermodynamic arrow for all (or most) such trajectories.

\subsection{Time symmetric boundary conditions}
Schulman\cite{Schulman1997} has considered the problem of universes
with two time boundary conditions.  Although the possibility of such
a universe remains questionable\cite{Zeh2005,Schulman2005b}, it will
be useful to consider such a situation here.  In these conditions
there is a requirement both that the universe begins in the special
initial region of state space $\Delta_0$, and at a remote future
time $\tau$ ends in the special final region of state space
$\Delta_0^T$.

A simple time reversal is not sufficient to deal with this.  The
possible trajectories of the system are those that pass through
$\phi^{(\tau)}(\Delta_0) \cap T(\Delta_0)$ at $t=\tau$.  Equivalent
conditions are $\Delta_0 \cap \phi^{-(\tau)} \circ T(\Delta_0)$ at
$t=0$ or $\phi^{(\tau/2)}(\Delta_0) \cap T \circ
\phi_{T\tau/2}^{(\tau/2)}(\Delta_0)$ at $t=\tau/2$.

%
%
%
Schulman argues that, provided the time span $\tau/2$ is much
greater than the complete thermalisation time $t_{th}$, then during
the epoch $0 < t < t_{th}$ the universe will be indistinguishable
from an entropy increasing universe, and during the epoch
$\tau-t_{th} < t < \tau $ the universe will be indistinguishable
from an entropy decreasing one.

\section{The Computation Arrow}\label{s:trevcomp}
A physical computation is a physical embodiment of a combination of
logical operations.  A logical operation is an abstract
mathematical operation which takes a finite number of distinct input
logical states, $\{\alpha\}$, and maps them to a finite number of distinct output logical states, $\{\beta\}$.
Conventionally the input logical state uniquely determines the
output logical state, but there may be many input states
corresponding to the same output state.  If this is the case, the
operation is called logically irreversible\cite{Lan61}.
\begin{quote}
We shall call a device logically irreversible if the output of a
device does not uniquely define the inputs.
\end{quote}
If each $\beta$ output state has only one possible $\alpha$ input
state, then the operation is called logically reversible.

The basic operations we need to consider are the NOT operation and
the RESET TO ZERO (RTZ) operations (see Tables \ref{tb:not} and
\ref{tb:rtz})\footnote{For completeness we include the identity or
DO NOTHING operation, IDN, in Table \ref{tb:idn}.}. The RTZ
operation is perhaps less familiar than logical operations such as
AND, OR.  Nevertheless, all standard logical operations can be built
from suitable combinations of these two operations, and they are the
most widely studied logical operations from the point of view of
thermodynamics.

\begin{table}[htbp]
  \centering
  \parbox{0.3\textwidth}
  {
    \centering
    \begin{tabular}{c||c}
        \multicolumn{2}{c}{$NOT$} \\ \hline
        IN & OUT \\
        \hline \hline 0 & 1 \\ 1 & 0
    \end{tabular}
    \caption{Logical NOT\label{tb:not}}
  }
   \parbox{0.3\textwidth}
  {
    \centering
    \begin{tabular}{c||c}
        \multicolumn{2}{c}{$RTZ$} \\ \hline
        IN & OUT \\
        \hline \hline 0 & 0 \\ 1 & 0
    \end{tabular}
    \caption{Reset to Zero\label{tb:rtz}}
  }
 \parbox{0.3\textwidth}
  {
    \centering
    \begin{tabular}{c||c}
        \multicolumn{2}{c}{$IDN$} \\ \hline
        IN & OUT \\
        \hline \hline 0 & 0 \\ 1 & 1
    \end{tabular}
    \caption{Logical Identity\label{tb:idn}}
  }
\end{table}

The physical embodiment of a logical operation is a physical
process, that starts with the system in one of a finite number of
distinct regions of state space and evolves the system into one of a
finite number of distinct regions of state space.  The distinct
regions of state space represent the input and output logical
states.  The same region can (and often will) represent both an
input and an output state.  The process embodies the logical
operation precisely when states in the region of state space
corresponding to an input logical state always end in the region of
state space corresponding to the output logical state that results
from the action of the logical operation upon that input logical
state.

To analyse the thermodynamics, take a state space $\Omega=\Omega_S
\otimes \Omega_E$, which is the product of the logical processing
system $\Omega_S$ and environment $\Omega_E$ state spaces.  In an
entropy increasing universe, we assume the environment is initially
in some region $E_0 \subset \Omega_E$ and there are no correlations
with the system. Each input logical state $\alpha$ is represented by a
region of the state space of the system $A_\alpha \subset
\Omega_S$, such that $A_\alpha \cap A_{\alpha^\prime} = \emptyset$ for
$\alpha \neq \alpha^\prime$. Similarly each output logical state $\beta$ is represented by a
region of the state space of the system $A_\beta \subset
\Omega_S$, such that $A_\beta \cap A_{\beta^\prime} = \emptyset$ for
$\beta \neq \beta^\prime$. It is usually the case, and we will assume it
here, that the input and output states of a logical operation are
time reversal invariant subspaces: $A_\alpha^T=A_\alpha$ and
$A_\beta^T=A_\beta$.

If the logical operation $L$ maps logical states $\alpha
\stackrel{L}{\rightarrow} \beta$, then the dynamic map
$\phi_L^{(t_L)}$, acting over the duration $t_L$, embodies that
operation if, and only if, $\forall \alpha \stackrel{L}{\rightarrow}
\beta$
\begin{equation}
\phi_L^{(t_L)}(A_\alpha \otimes E_0) \subseteq A_\beta \otimes
\Omega_E
\end{equation}

At the start of the physical operation, the system will be in one of the regions that represents one of the input logical states
\begin{equation}
\Delta_0=\cup_\alpha A_\alpha \otimes E_0
\end{equation}
At the end of the physical operation, the system and environment
will be located in the region:
\begin{equation}
\Delta_{t_L}=\phi_L^{(t_L)}(\Delta_0)=\cup_\alpha
\phi_L^{(t_L)}(A_\alpha \otimes E_0)\subseteq \cup_\beta A_\beta
\otimes \Omega_E
\end{equation}
In an entropy increasing universe, we assume that microscopic
correlations between the system and the environment play no future
role.  If we are not considering time reversals, therefore, for
future evolutions of the system we can replace $\Delta_{t_L}$ with the coarse grained region:
\begin{equation}
\overline{\Delta}_{t_L}= \cup_\beta A_\beta \otimes \overline{E}_{t_L}
\end{equation}
where $\forall E^\prime \subseteq \Omega_E$ such that $\Delta_{t_L}
\subseteq \cup_\beta A_\beta \otimes E^\prime$, then
\begin{equation} \Delta_{t_L} \subseteq \cup_\beta A_\beta \otimes
\overline{E}_{t_L} \subseteq \cup_\beta A_\beta \otimes E^\prime
\end{equation}

\subsection{Temporal reversal}\label{ss:temprev}
We will now need to consider how the physical embodiment of a logical operation is affected by a temporal reversal.  We will find there is an apparent temporal asymmetry in some of these physical processes.

The temporal reversal of the physical operation, at time
$\frac{1}{2}t_L$, involves the system and environment starting in
the region of state space $\Delta_{t_L}^T=T(\Delta_{t_L})$, with the evolution
$\phi_{TL}^{(t)}(\Delta)=T \circ \phi_L^{(t_L-t)}\circ
\phi_L^{-(t_L)}\circ T (\Delta)$.  This gives $\phi_{TL}^{(t_L)}(\Delta)=T \circ \phi_L^{-(t_L)}\circ T (\Delta)$ for the complete operation, acting over $0 \leq t \leq t_L$.

By definition
\begin{equation}
T \circ \phi_L^{(t_L)}(A_\alpha \otimes E_0)\subseteq A_\beta
\otimes \Omega_E
\end{equation}
and
\begin{equation}
\phi_{TL}^{(t_L)}\left(T \circ \phi_L^{(t_L)}(A_\alpha \otimes
E_0)\right)=A_\alpha \otimes E_0^T \subseteq A_\alpha \otimes
\Omega_E
\end{equation}

$\phi_{TL}^{(t_L)}$ has acted as a map from the system being in one
of the regions of state space corresponding to a logical state
$\beta$ to being in a region of state space corresponding to the initial
logical state $\alpha$.  This is what we expect from a temporal reversal.

If the operation was logically reversible, then for each $\beta$ there was only one $\alpha$ for which $A_\alpha$ was mapped to $A_\beta$.  $A_\beta$ is then mapped back to $A_\alpha$ by $\phi_{TL}^{(t_L)}$.  The temporal reversal operation acts upon each $A_\beta$ and mapping it to a specific $A_\alpha$.  It is embodying the inverse logical operation to $L$.

Logically irreversible
operations are not invertible functions.  There will be more that one region $A_\alpha$ which was mapped to a given
$A_\beta$ by the operation $L$.  The effect of the map $\phi_{TL}^{(t_L)}$ on the region $A_\beta$ will be to map it into
several different $A_\alpha$ regions.  Where $L$ is a logically irreversible operation, $\phi_{TL}^{(t_L)}$ does not appear to be the physical embodiment of a logical operation.  There appears to be a fundamental temporal asymmetry in the physical embodiments of logical operations.

\subsection{Indeterministic operations}\label{ss:indetop}
To better understand the time reversal of
logically irreversible operations, we need to widen the class of
operations we are considering, to include {\em
indeterministic}\footnote{While indeterministic operations can be
well defined, and can always be embodied by physical processes, it has been
argued that indeterministic operations do not count as logical
operations(\cite{SLGP05}, for example), although indeterministic operations are
required for computational complexity classes such as $BPP$, and so
form a part of computational logic.  As this point is not important
for the discussion here, we will reserve `logical operation' for
logically deterministic operations in this paper, and refer to
logically indeterministic operations as simply `indeterministic
operations'.} operations\cite{Mar05b}:
\begin{quote}
We shall call a device logically indeterministic if the input to a
device does not uniquely define the outputs.
\end{quote}

The time reversal of the logically reversible $IDN$ and the $NOT$
operations result in the $IDN$ and $NOT$ operations, respectively.
Time reversal of logically irreversible $RTZ$, however, results in
the indeterministic operation Unset From Zero ($UFZ$) in Table
\ref{tb:ufz}.  Note that the operation $UFZ$ does fulfil the
requirement of logical reversibility, above.  For completeness, we
also add the indeterministic, irreversible operation Randomise
($RND$) in Table \ref{tb:rnd}.
\begin{table}[htbp]
  \centering
  \parbox{0.4\textwidth}
  {
    \centering
    \begin{tabular}{c||c}
        \multicolumn{2}{c}{$UFZ$} \\ \hline
        IN & OUT \\
        \hline \hline 0 & 0 \\ 0 & 1
    \end{tabular}
    \caption{Unset From Zero\label{tb:ufz}}
  }
   \parbox{0.4\textwidth}
  {
    \centering
    \begin{tabular}{c||c}
        \multicolumn{2}{c}{$RND$} \\ \hline
        IN & OUT \\
        \hline \hline 0 & 0 \\ 0 & 1 \\ 1 & 0 \\ 1 & 1
    \end{tabular}
    \caption{Randomise\label{tb:rnd}}
  }
\end{table}

A computation is not simply a sequence of operations. It is an
ordered sequence of particular logical operations.  If a Universal
Turing Machine is constructed out of a collection of physical
processes implementing a particular set of logically deterministic
operations, the time reversal of those physical processes certainly
does not produce the same set of operations.  If the Universal
Turing Machine was constructed using deterministic, logically
irreversible operations, the time reversal would not include any
logically irreversible operations but would include indeterministic
operations.  This would not be a Universal Turing Machine.

Logically irreversible operations may be simulated by logically
reversible operations, but under time reversal this still does not
recover the original computation.  The logically reversible
simulation of the $RTZ$ operation is given in Table \ref{tb:revrtz},
and its time reversal in Table \ref{tb:revufz}.
\begin{table}[htbp]
  \centering
  \parbox{0.4\textwidth}
  {
    \centering
    \begin{tabular}{cc||cc}
        \multicolumn{2}{c}{IN} & \multicolumn{2}{c}{OUT} \\
        \hline \hline 0 & 0 & 0 & 0 \\ 1 & 1 & 1 & 0
    \end{tabular}
    \caption{Simulating $RTZ$\label{tb:revrtz}}
  }
   \parbox{0.4\textwidth}
  {
    \centering
    \begin{tabular}{cc||cc}
        \multicolumn{2}{c}{IN} & \multicolumn{2}{c}{OUT} \\
        \hline \hline 0 & 0 & 0 & 0\\ 1 & 0 & 1 & 1
    \end{tabular}
    \caption{Simulating $UFZ$\label{tb:revufz}}
  }
\end{table}

The time reversal is not a reversible simulation of $RTZ$, it is a
deterministic simulation of $UFZ$.  Although, in this case, both
simulations can be achieved by the same logical operation (the
$CNOT$ gate), the particular operation that is being simulated
changes. A sequence of operations simulating irreversible operations
becomes a sequence of operations simulating indeterministic
operations.  If the Universal Turing Machine was constructed using
deterministic, logically reversible operations, simulating logically
irreversible operations, the time reversal would not include any
simulations of logically irreversible operations but would include
simulations of indeterministic operations.  This would still not be
a Universal Turing Machine.  The time reversal of a Universal Turing
Machine is not a Universal Turing Machine.  So it would appear that
a computation, as a physical process, may have an arrow of time.

\subsection{Logical reversal}
With the concept of indeterministic operations in place, we can define a new concept, the logical reversal of an operation.
A logical operation, $L$, has the
logical reversal operation, $RL$, which has the same mapping on the logical states,
as the temporal reversal of a physical implementation of the original operation $L$.  If the operation $L$ maps the input state $\alpha$ to the output state $\beta$, then the logical reversal maps $\beta$, as an input state to $\alpha$, as an output state.  When $L$ is a logically irreversible operation, there are many $\alpha$ mapped to a given $\beta$, so $RL$ will be a logically indeterministic operation, as it may map $\beta$ to one of many $\alpha$.  Similarly, if $L$ is a logically indeterministic operation, then $RL$ will be a logically reversible operation.  Unlike the temporal reversal, the logical reversal acts in the same time direction as the original operation.

We construct $RL$, first by defining proportions (according to a measure $\mu$ on the state space)
of the states acted up by the operation $L$, that start and end in the regions representing the logical states.  The system starts in the region $\Delta_0=\cup_\alpha A_\alpha \otimes E_0$ and ends in the region $\Delta_{t_L}=\phi_L^{(t_L)}(\cup_\alpha A_\alpha \otimes E_0)$.
\begin{enumerate}
\item That start in logical state $\alpha$
\[
W_L(\alpha) = \frac{\mu\left((A_\alpha \otimes \Omega_E)\cap
\Delta_0\right)}{\mu\left(\Delta_0\right)}
\]
\item That end in logical state $\beta$ given they started in
$\alpha$
\[
 W_L(\beta|\alpha) =
\frac{\mu\left((A_\beta \otimes \Omega_E)\cap
\phi_L^{(t_L)}(A_\alpha \otimes \Omega_E)\cap
\Delta_{t_L}\right)}{\mu\left((A_\alpha \otimes \Omega_E)\cap
\Delta_0\right)} \]
\item That start in logical state
$\alpha$ and end in logical state $\beta$
\[
 W_L(\alpha,\beta) =
\frac{\mu\left((A_\beta \otimes \Omega_E)\cap
\phi_L^{(t_L)}(A_\alpha \otimes \Omega_E)\cap
\Delta_{t_L}\right)}{\mu\left(\Delta_0\right)}
\]
\item That end in logical state $\beta$
\[
W_L(\beta) = \frac{\mu\left((A_\beta \otimes \Omega_E)\cap
\Delta_{t_L}\right)}{\mu\left(\Delta_0\right)}
\]
\item That started in logical state $\alpha$, given that they ended in
logical state $\beta$
\[
W_L(\alpha|\beta)=\frac{\mu\left((A_\beta \otimes \Omega_E)\cap
\phi_L^{(t_L)}(A_\alpha \otimes \Omega_E)\cap
\Delta_{t_L}\right)}{\mu\left((A_\beta \otimes \Omega_E)\cap
\Delta_{t_L}\right)}
\]
\end{enumerate}
These proportions satisfy the expected relationship $W_L(\alpha,\beta)=W_L(\beta|\alpha)W_L(\alpha)=W_L(\alpha|\beta)W_L(\beta)$.
For logically deterministic operations
\[
W_L(\alpha|\beta) \in \{0,1\}
\]
while for logically reversible operations
\[
W_L(\beta|\alpha) \in \{0,1\}
\]
We do not include input or output states with measure zero, so
$W_L(\alpha)\neq 0$ and $W_L(\beta) \neq 0$.  If
$W_L(\alpha|\beta)=0$ for the measure $\mu$, it will be zero for all
other measures, absolutely continuous with $\mu$, that are preserved
by the dynamics.  Equivalent statements also hold true for
$W_L(\alpha|\beta)=1$, $W_L(\beta|\alpha)=0$ and
$W_L(\beta|\alpha)=1$.

Now consider the temporal reversal $TL$ of this physical process.  This acts upon logical states $\beta$ and produces logical states $\alpha$, with proportions
\begin{eqnarray*}
W_{TL}(\beta) &=& \frac{\mu\left((A_\beta \otimes \Omega_E)\cap
\Delta_{t_L}\right)}{\mu\left(\Delta_{t_L}\right)} \\
W_{TL}(\alpha|\beta)&=&\frac{\mu\left(\phi_{TL}^{(t_L)}(A_\beta
\otimes \Omega_E)\cap (A_\alpha \otimes \Omega_E)\cap
\Delta_0\right)}{\mu\left((A_\beta \otimes \Omega_E)\cap \Delta_{t_L}\right)}\\
W_{TL}(\alpha,\beta) &=& \frac{\mu\left(\phi_{TL}^{(t_L)}(A_\beta
\otimes \Omega_E)\cap (A_\alpha \otimes \Omega_E)\cap
\Delta_0\right)}{\mu\left(\Delta_{t_L}\right)}\\
W_{TL}(\alpha) &=& \frac{\mu\left((A_\alpha \otimes \Omega_E)\cap
\Delta_0\right)}{\mu\left(\Delta_{t_L}\right)}\\
W_{TL}(\beta|\alpha)& =& \frac{\mu\left(\phi_{TL}^{(t_L)}(A_\beta
\otimes \Omega_E)\cap (A_\alpha \otimes \Omega_E)\cap
\Delta_0\right)}{\mu\left((A_\alpha \otimes \Omega_E)\cap
\Delta_0\right)}
\end{eqnarray*}
It is straightforward to show that as
\[
W_{TL}(\beta)=W_{L}(\beta)
\]
then
\[
W_{TL}(\alpha)=W_{L}(\alpha)
\]
and
\[
W_{TL}(\beta|\alpha)=W_{L}(\beta|\alpha)
\]
It is also clear, by definition, that the temporal reversal of $TL$
is just $L$:
\[
TTL\equiv L
\]

We will now define the logical reversal operation of $L$, as a map
from the set of logical states $\{\beta\}$ to the set of logical states $\{\alpha\}$,
\[
\{\beta\} \stackrel{RL}{\rightarrow} \{\alpha\}
\]
in the \textit{same} time direction as $L$, with a dynamic map
$\phi_{RL}^{(t_L)}$ such that
$W_{RL}(\alpha|\beta)=W_{TL}(\alpha|\beta)=W_{L}(\alpha|\beta)$.  The system is initially in the region $\Lambda_0=\cup_\beta
A_\beta \otimes E_0$ and ends in the region $\Lambda_{t_L}
=\phi_{RL}^{(t_L)}(\Lambda_0)$.  The map should satisfy:
\begin{eqnarray*}
W_{RL}(\beta) &=& \frac{\mu\left((A_\beta \otimes \Omega_E)\cap
\Lambda_0\right)}{\mu\left(\Lambda_0\right)} \\
W_{RL}(\alpha|\beta)&=&\frac{\mu\left(\phi_{RL}^{(t_L)}(A_\beta
\otimes \Omega_E)\cap (A_\alpha \otimes \Omega_E)\cap
\Lambda_{t_L}\right)}{\mu\left((A_\beta \otimes \Omega_E)\cap \Lambda_0\right)}\\
W_{RL}(\alpha,\beta) &=& \frac{\mu\left(\phi_{RL}^{(t_L)}(A_\beta
\otimes \Omega_E)\cap (A_\alpha \otimes \Omega_E)\cap
\Lambda_{t_L}\right)}{\mu\left(\Lambda_0\right)}\\
W_{RL}(\alpha) &=& \frac{\mu\left((A_\alpha \otimes \Omega_E)\cap
\Lambda_{t_L}\right)}{\mu\left(\Lambda_0\right)}\\
W_{RL}(\beta|\alpha)& =& \frac{\mu\left(\phi_{RL}^{(t_L)}(A_\beta
\otimes \Omega_E)\cap (A_\alpha \otimes \Omega_E)\cap
\Lambda_{t_L}\right)}{\mu\left((A_\alpha \otimes \Omega_E)\cap
\Lambda_{t_L}\right)}
\end{eqnarray*}
Again, it is straightforward that
\[
W_{RL}(\beta)=W_{L}(\beta)
\]
leads to
\[
W_{RL}(\alpha)=W_{L}(\alpha)
\]
and
\[
W_{RL}(\beta|\alpha)=W_{L}(\beta|\alpha)
\]
By definition \[ RRL \equiv L\]

There is a straightforward method for constructing $\phi_{RL}$:
\begin{enumerate}
\item Partition each $\beta$ region into $(\alpha,\beta)$
subregions, $A_\beta=\cup_\alpha A_{(\alpha|\beta)}$, with
$A_{(\alpha|\beta)}\cap A_{(\alpha^\prime|\beta)}=\emptyset$ ,
$\alpha \neq \alpha^\prime$ such that
\[\frac{\mu\left(\alpha,\beta\right)}{\mu\left( \beta
\right)}=W_{L}(\alpha|\beta)\]
\item The evolution of the system must prevent transitions between the
subregions
\[
\phi(A_{(\alpha|\beta)})\cap \phi(A_{(\alpha^\prime|\beta^\prime)})
= \emptyset \;\; \forall \alpha \neq \alpha^\prime,  \beta \neq
\beta^\prime
\]
\item Define regions $A^\prime_\alpha$ by joining the $\alpha$ subregions together, from different $\beta$
regions
\[
A^\prime_\alpha=\cup_\beta A_{(\alpha|\beta)}
\]
and remove barriers to transitions between subregions with the same
$\alpha$ value.
\item Evolve the distinct $\alpha$ regions to their final location in state
space:
\[
\phi(A^\prime_\alpha)\subseteq A_\alpha
\]
\end{enumerate}
Further refinements are necessary for thermodynamic optimisation.
 Explicit physical processes by which the operations $UFZ$ and $RND$
can be constructed and optimised are given in \cite{Mar05b} and for
generic operations in \cite{Maroney2007b,Turgut2006}.

\subsection{Computational reversal}\label{ss:trevseq}
We will now consider sequences of operations, in a normal entropy
increasing universe.  We will not specify the particular set of
operations.  Our objective is not to consider the properties of a
particular sequence of logical operations, or even of any sequence
of logical operations intended for a particular purpose.  We wish to
consider the properties of \textit{any} process that can be defined
exclusively in terms of logical operations acting upon sets of
logical states.

In this general situation, we will start with a set of logical states
$\{\alpha_0\}$.  This is acted on by some logical operation $L_0$,
and mapped to the set of output states $\{\alpha_1\}$.  As we are in an
entropy increasing universe, we may assume that any microscopic
correlations that have developed between the information processing
apparatus and the environment play no role in the future evolution
of the system.  The next logical operation $L_1$ then maps the logical states
$\{\alpha_1\}$ to the logical states $\{\alpha_2\}$, and so on.

This leads to a time ordered sequence of logical operations, and the sequences of logical states which can occur as a result of these operations:
\[
\{\alpha_0\} \stackrel{L_1}{\rightarrow} \{\alpha_1\}
\stackrel{L_2}{\rightarrow} \dots \stackrel{L_i}{\rightarrow}
\{\alpha_i\} \stackrel{L_{i+1}}{\rightarrow} \dots
\stackrel{L_f}{\rightarrow} \{\alpha_f\}
\]
We will refer to this sequence as $S_1\{L_i\}$.

We now consider the entropy decreasing universe that results from a time reversal.  In the time reversed, entropy decreasing universe, this computational sequence becomes
$S_2\{TL_i\}$:
\[
\{\alpha_f\} \stackrel{TL_f}{\rightarrow} \{\alpha_{f-1}\}
\stackrel{TL_{f-1}}{\rightarrow} \dots
\stackrel{TL_{i+1}}{\rightarrow} \{\alpha_i\}
\stackrel{TL_i}{\rightarrow} \dots \stackrel{TL_1}{\rightarrow}
\{\alpha_0\}
\]
As we have noted in Sections \ref{ss:temprev} and \ref{ss:indetop}, the sequence of operations $S_2\{TL_i\}$, involving
the time reversed $TL$ operations, will not, in general, resemble
the same computational process as $S_1\{L_i\}$.  If $S_1\{L_i\}$ is representing the logical operations performed by a Turing machine, there is no guarantee that the sequence of operations $S_2\{TL_i\}$ resembles a computation at all.

We now return to the original entropy increasing condition, and construct a physical system in an entropy increasing universe that implements the operations $\{RL_i\}$, the logical reversals of the operations $\{L_i\}$.   Staring with initial logical states $\{\alpha_f\}$, a measure
$\mu$ such that the physical representation of the states have
weights $W_{RL_f}(\alpha_f)=W_{L_f}(\alpha_f)$, and the reversal
operations $\{RL_i\}$, such that
$W_{RL_i}(\alpha_{i-1}|\alpha_i)=W_{L_i}(\alpha_{i-1}|\alpha_i)$, this leads to the time ordered computational sequence $S_3\{RL_i\}$:
\[
\{\alpha_f\} \stackrel{RL_f}{\rightarrow} \{\alpha_{f-1}\}
\stackrel{RL_{f-1}}{\rightarrow} \dots
\stackrel{RL_{i+1}}{\rightarrow} \{\alpha_i\}
\stackrel{RL_i}{\rightarrow} \dots \stackrel{RL_1}{\rightarrow}
\{\alpha_0\}
\]
Again, this sequence is quite distinct from $S_1\{L_i\}$.  As a sequence of operations, acting upon the logical states, it is identical to $S_2\{TL_i\}$, as $RL_i$, by definition, implements the same operation from the set of states $\{\alpha_i\}$ to $\{\alpha_{i-1}\}$ as does $TL_i$.  However, as noted $S_2\{TL_i\}$ need not resemble a computation and so neither need $S_3\{RL_i\}$.

Now we complete the central argument of the paper.  The time reversal of the universe containing the sequence
$S_3\{RL_i\}$, gives the sequence $S_4\{TRL_i\}$:
\[
\{\alpha_0\} \stackrel{TRL_1}{\rightarrow} \{\alpha_1\}
\stackrel{TRL_2}{\rightarrow} \dots \stackrel{TRL_i}{\rightarrow}
\{\alpha_i\} \stackrel{TRL_{i+1}}{\rightarrow} \dots
\stackrel{TRL_f}{\rightarrow} \{\alpha_f\}
\]
This is now in an entropy decreasing universe.
However, it follows straightforwardly from the definitions above, that $TRL_i \equiv
RTL_i \equiv L_i$, so $S_4\{TRL_i\}$ is
\[
\{\alpha_0\} \stackrel{L_1}{\rightarrow} \{\alpha_1\}
\stackrel{L_2}{\rightarrow} \dots \stackrel{L_i}{\rightarrow}
\{\alpha_i\} \stackrel{L_{i+1}}{\rightarrow} \dots
\stackrel{L_f}{\rightarrow} \{\alpha_f\}
\]
The time ordered sequence $S_4\{TRL_i\}$ is exactly the same set of logical
operations as $S_1\{L_i\}$, performed in the same order, and on the
same set of logical states.  $S_4\{TRL_i\}$ takes place in an
entropy decreasing universe.

For any computational process consisting of a sequence of logical
operations on a set of logical states, in an entropy increasing
universe, the same computational process is possible in an entropy
decreasing universe.  Although we were able to conclude in Section
\ref{ss:indetop}, above, that computational processes may have an
intrinsic arrow, it does not appear to be the case that this arrow
must be aligned with the thermodynamic arrow.

\subsection{Landauer's Principle}\label{s:landp} Landauer's Principle
is used as the basis for almost all conclusions regarding the
thermodynamic properties of physical computation, yet the conclusion
of the previous section seems to run counter to many widespread
statements of this Principle:
\begin{quote}
To erase a bit of information in an environment at temperature $T$
requires dissipation of energy $\geq kT \ln 2$. \cite{Cav90,Cav93}

in erasing one bit \ldots of information one dissipates, on average,
at least $k_B T \ln\left(2\right)$ of energy into the environment.
\cite{Pie00}

a logically irreversible operation must be implemented by a
physically irreversible device, which dissipates heat into the
environment \cite{Bub02}

erasure of one bit of information increases the entropy of the
environment by at least $k \ln 2$ \cite{LR03}[pg 27]

any logically irreversible manipulation of data $\ldots$ must be
accompanied by a corresponding entropy increase in the
non-information bearing degrees of freedom of the information
processing apparatus or its environment.  Conversely, it is
generally accepted that any logically reversible transformation of
information can in principle be accomplished by an appropriate
physical mechanism operating in a thermodynamically reversible
fashion. \cite{Ben03}

Computations are accompanied by dissipation \ldots Landauer has
shown that computation requires irreversible processes and heat
generation.\cite{Schulman2005a}
\end{quote}
It is Landauer's Principle on which Schulman basis the alignment of
the thermodynamic and the computational arrows of time.

If Landauer's Principle is truly regarded as ``the basic principle
of the thermodynamics of information processing''\cite{Ben03}, how
does this reconcile with the argument of the previous Section, that
exactly the same information processing operations can take place in
an entropy decreasing, as an entropy increasing universe? Does the
computer act as a kind of Maxwell's Demon, dissipating heat against
overall the anti-entropic direction?

The answer is, straightforwardly, no.  As has been noted many times
before\cite{EN99,Mar02,Nor05}, Landauer's Principle is not really a
principle.  It is a theorem, of statistical mechanics,
derived\cite{Pie00,Turgut2006,SLGP05,Maroney2007b} on the assumption
that the computation is taking place in an entropy increasing
universe. All justifications of Landauer's Principle, from
\cite{Lan61} onwards, make this assumption.  We will briefly review
the derivation of Landauer's Principle in an entropy increasing
universe, to see how the derivation turns out in an entropy
decreasing universe.

\subsubsection{Entropy increase} The states of the
physical system embodying logical state $\alpha$ will be represented
by density matrix $\rho_\alpha$, and $\beta$ by $\rho_\beta$.  We
assume\footnote{This is normal practice in the thermodynamics of
computation. In \cite{Maroney2007b} this assumption is relaxed. The essential conclusions of this Section are
not affected.} that the input logical states $\{\alpha\}$ and output
logical states $\{\beta\}$ are represented by states of physical
systems with the same entropy $S$ and mean energies $U$, so that
$\forall \alpha, \beta$:
\begin{eqnarray}
S&=-k\trace{\rho_\alpha \ln \func{\rho_\alpha}}=&-k\trace{\rho_\beta
\ln\func{\rho_\beta}}\\
U&=\trace{H_S \rho_\alpha}=&\trace{H_S \rho_\beta}
\end{eqnarray}
The input logical states occur with probability $P_\alpha$, and the
logical operation is defined by the probabilities $P(\beta|\alpha)$.

In an entropy increasing universe, we make the following
assumptions:
\begin{enumerate}
\item The evolution of the system and environment is described by Hamiltonian
dynamics, composed of internal energies of the system $H_S$ and
environment $H_E$, together with an interaction potential $V_{SE}$:
\[
H=H_S \otimes I_E +I_S \otimes H_E +V_{SE}
\]
\item The environment is initially in a Gibbs canonical state, at some temperature $T$, and
there are no initial correlations between the system and the
environment.
\begin{eqnarray}
\rho_E(T)&=& \frac{e^{-H_E/kT}}{\trace{e^{-H_E/kT}}} \\
\rho_0&=&\sum_\alpha P(\alpha)\rho_\alpha \otimes \rho_E(T)
\end{eqnarray}

\item The interaction energy between system and environment is
negligible both before
\[
\trace{V_{SE}\rho_0} \approx 0
\]
and after
\[
\trace{V_{SE}e^{-\imath H t}\rho_0 e^{\imath H t}} \approx 0
\]
the interaction.
\end{enumerate}

For the Hamiltonian $H$ to embody the logical operation:
\[
\partrace{e^{-\imath H t}\rho_\alpha \otimes \rho_E(T)e^{\imath
H t}}{E}=\sum_\beta P(\beta|\alpha)\rho_\beta
\]
 It is a well known
calculation\cite{Gib1902,Tol1938,Par89a,Pie00,Maroney2007b} to show,
using:
\begin{eqnarray*}
\rho_I&=&\sum_\alpha P(\alpha) \rho_\alpha \\
\rho_t&=&e^{-\imath H t}\rho_0 e^{\imath H t} \\
P(\beta)&=&\sum_\alpha P(\beta|\alpha)P(\alpha) \\
\rho_F&=&\partrace{\rho_t}{E}=\sum_\beta P(\beta) \rho_\beta \\
\rho^\prime_E&=&\partrace{\rho_t}{S}
\end{eqnarray*}
that two inequalities follow:
\begin{eqnarray}
\trace{\rho_I \ln \func{\rho_I}}+\trace{\rho_E(T) \ln
\func{\rho_E(T)}} &\geq &\trace{\rho_F \ln
\func{\rho_F}}+\trace{\rho^\prime_E \ln \func{\rho^\prime_E}} \\
\trace{\rho^\prime_E\left(\ln
\func{\rho^\prime_E}+\frac{H_E}{kT}\right)}
&\geq&\trace{\rho_E(T)\left(\ln
\func{\rho_E(T)}+\frac{H_E}{kT}\right)}
\end{eqnarray}
which combine to give
\begin{equation}
\sum_\alpha P(\alpha)\ln P(\alpha)-\sum_\beta P(\beta)\ln P(\beta)
\geq \frac{\trace{H_E \rho_E(T)}}{kT} - \frac{\trace{H_E
\rho^\prime_E}}{kT}
\end{equation}
This yields the standard form of Landauer's Principle, in an entropy
increasing universe:
\[
\Delta Q\geq -\Delta H k T \ln (2)
\]
where $\Delta Q$ is the expectation value for the heat generated in
an environment at temperature $T$ and $\Delta H$ is the change in
Shannon information over the course of the operation
\[
\Delta H=\sum_\alpha P(\alpha)\log_2 P(\alpha)-\sum_\beta
P(\beta)\log_2 P(\beta)
\]

For logically deterministic, reversible computations, it is always
the case that $\Delta H=0$.  These operations do not need to
generate heat. On the other hand, for logically deterministic,
irreversible operations $\Delta H<0$ and so the heat generated in
the environment is always positive.  This is the basis of the claim
that logically irreversible operations must be entropy
increasing\footnote{In \cite{Mar02,Mar05b,Maroney2007b} it is argued
that even this heat generation is not necessarily thermodynamically
irreversible.}.

\subsubsection{Entropy decrease} In an entropy decreasing universe,
we would still make the assumptions that the input logical states
$\{\alpha\}$ and output logical states $\{\beta\}$ are represented
by physical systems with the same entropy and mean energies. The
logical state $\alpha$ is represented by the density matrix
$\rho_\alpha$, and $\beta$ by $\rho_\beta$, as before.  The input
logical states occur with probability $P_\alpha$, and the logical
operation is defined by the probabilities $P(\beta|\alpha)$.

We continue to assume:
\begin{enumerate}
\item The evolution of the system and environment is described by Hamiltonian dynamics.
\[
H^\prime=H^\prime_S \otimes I_E +I_S \otimes H^\prime_E
+V^\prime_{SE}
\]
\item The interaction energy between system and environment is
negligible both before
\[
\trace{V^\prime_{SE}\rho_0} \approx 0
\]
and after
\[
\trace{V^\prime_{SE}e^{-\imath H^\prime t}\rho_0 e^{\imath H^\prime
t}} \approx 0
\]
the interaction.
\end{enumerate}
but the imposition of a future boundary condition must require the
local conditions to be:
\begin{enumerate}\setcounter{enumi}{2}
\item {\em After} the operation the environment is in a Gibbs canonical state, at some temperature $T$, and
there are no final microscopic correlations between the system and
the environment.
\end{enumerate}
Now, for the Hamiltonian $H^\prime$ to fulfil these conditions and
embody the logical operation it is necessary that
\[
\partrace{e^{\imath H^\prime t}\rho_\beta \otimes \rho_E(T)e^{-\imath
H^\prime t}}{E}=\sum_\alpha \frac{P(\beta|\alpha)P(\alpha)}
{\sum_{\alpha^\prime}
P(\beta|\alpha^\prime)P(\alpha^\prime)}\rho_{\alpha}
\]
and
\[
\rho_t=\sum_{\beta,\alpha} P(\beta|\alpha)P(\alpha)\rho_\beta
\otimes \rho_E(T)
\]

Using:
\begin{eqnarray*}
\rho_0&=&e^{\imath H^\prime t}\rho_t e^{-\imath H^\prime t} \\
\rho_I&=&\partrace{\rho_0}{E}=\sum_\alpha P(\alpha) \rho_\alpha \\
\rho^\prime_E&=&\partrace{\rho_0}{S}\\
P(\beta)&=&\sum_\alpha P(\beta|\alpha)P(\alpha) \\
\rho_F&=&\partrace{\rho_t}{E}=\sum_\beta P(\beta) \rho_\beta
\end{eqnarray*}
the two inequalities become
\begin{eqnarray*}
\trace{\rho_F \ln \func{\rho_F}}+\trace{\rho_E(T) \ln
\func{\rho_E(T)}} &\geq &\trace{\rho_I \ln
\func{\rho_I}}+\trace{\rho^\prime_E \ln \func{\rho^\prime_E}} \\
\trace{\rho^\prime_E\left(\ln
\func{\rho^\prime_E}+\frac{H^\prime_E}{kT}\right)}
&\geq&\trace{\rho_E(T)\left(\ln
\func{\rho_E(T)}+\frac{H^\prime_E}{kT}\right)}
\end{eqnarray*}
which combine to give
\[
-\sum_\alpha P(\alpha)\ln P(\alpha)+\sum_\beta P(\beta)\ln P(\beta)
\geq \frac{\trace{H^\prime_E \rho_E(T)}}{kT} -
\frac{\trace{H^\prime_E \rho^\prime_E}}{kT}
\]

Paying careful attention to the fact that $\rho_E(T)$ is now the
{\em final} state of the environment the statistical mechanical
calculation leads to:
\[
\Delta Q \leq -\Delta H k T \ln (2)
\]
where $\Delta Q$ is the expectation value for the heat generated in
an environment.

For logically deterministic, irreversible operations $\Delta H<0$
and so the heat generated in the environment is {\em less than} the
positive number $-\Delta H k T \ln (2)$.  For logically
deterministic, reversible computations, $\Delta H=0$ as before, but
this now just means the heat generation must be {\em less than}
zero. In an entropy decreasing universe, the derivation of
Landauer's Principle yields a {\em maximum} heat generation.  If
less than the maximum heat is generated, then there will have been
an uncompensated decrease in the entropy of the universe.

This is, of course, exactly what we should have expected!  In
entropy decreasing universes, the physical processes which embody
computations are, generically, entropy decreasing processes. There
is no contradiction between the statistical mechanical basis of
Landauer's Principle, and the conclusions of Section
\ref{ss:trevseq}.

\section{The Correlation Arrow}\label{s:corr}
It has been argued in the previous Sections that, although a
computer may possess a computational arrow, its functioning as a
physical process does not imply the alignment of that arrow with the
thermodynamic arrow.  The argument was based upon all the same
computational operations that can take place in an entropy
increasing universe being physically possible in an entropy
decreasing universe.  This still leaves open the possibility that it
is much more likely for systems to develop which process information
in the same direction as entropy increase, than systems which
process information in the direction of entropy decrease.

Turning to this question, the arguments will seem less concrete than
in the previous sections. This is a consequence of the need to
consider if cosmological boundary conditions, over the lifetime of
the universe, on the state of the whole universe, may have
influences on the localised behaviour of systems, operating over
short timescales, at a time in between, and very far from, either
initial or final state of the universe.  It is unclear how secure
the chain of reasoning involved in understanding such influences can
be (see \cite{Earman2006}, for example, for a sceptical view).

How might such an argument be constructed? Hawking\cite{Hawking1994}
suggests:
\begin{quote}
If one imposes a final boundary condition \ldots one can show that
the correlation between the computer memory and the surroundings is
greater at early times than at late times.  In other words, the
computer remembers the future, but not the past.
\end{quote}
Similar arguments are presented in \cite{Hawking1985,Hawking1987}.

The acquisition of information requires an increase in the
correlation between the computer and its surroundings.  A future
boundary condition, as interpreted in Section \ref{ss:entdec},
requires correlations to decrease in time.   To explore this
requires a move beyond the consideration of a computer as an
information processor.  We must take into account the nature of the
information that the system processes.  It is a system that acquires
new information about its surroundings and interacts with its
surroundings conditional upon the information it has acquired.  Such
behaviour has been characterised as an Information Gathering and
Utilising System, or $IGUS$.

\subsection{Information Gathering and Utilising Systems}
The behaviour of an $IGUS$ may be described as:
\begin{enumerate}
\item There is a correlation between the macroscopic states of the
internal states of an $IGUS$ and the macroscopic states of its
surroundings.
\item These macroscopic correlations occurred through an interaction
of the system with the surroundings, in the past.  At an earlier
point in time the macroscopic correlations did not exist. The
existing correlations are screened off by an earlier interaction.
\item New macroscopic correlations develop over time through conditional
interactions.  These can change the macroscopic internal states of
the system conditional upon the states of the surroundings, or
change the states of the surroundings, conditional upon the internal
states of the system.
\item Any macroscopic correlations between the current state of the
system and future states of its surroundings, are screened off by
the existing correlations and interactions between system and
environment that take place between the present and the future
time.\end{enumerate} The argument of Hawking is that such behaviour
is compatible with an initial boundary condition, but incompatible
with a future boundary condition.

We can examine this in two equivalent ways. The first is to consider
an $IGUS$ in an entropy increasing and in an entropy decreasing
universe. The second way is to consider the time reversal of these
two scenarios.  This will give a information processing system which
is the logical reversal of an $IGUS$, in an entropy decreasing and
in an entropy increasing universe, respectively.  We refer to the
logical reversal of an $IGUS$ as an $RIGUS$.  The statement that an
entropy decreasing universe is incompatible with the operation of an
$IGUS$ is equivalent to the statement that an entropy increasing
universe is incompatible with an $RIGUS$.

The question needing answering is whether an entropy increasing
universe prefers systems resembling an $IGUS$ over systems
resembling an $RIGUS$.  If so the same argument should support the
existence of an $RIGUS$ compared to an $IGUS$ in an entropy
decreasing universe.

The behaviour of an $RIGUS$ will appear as:
\begin{enumerate}
\item There is a correlation between the macroscopic states of the internal
states of an $RIGUS$ and the macroscopic states of its surroundings.
\item These macroscopic correlations will disappear through a conditional
interaction of the system with the surroundings, at some point in
future.  At a later point in time the macroscopic correlations will
not exist.
\item There decrease in macroscopic correlations over time is
through conditional interactions with the surroundings.    These can
change the macroscopic internal states of the system conditional
upon the states of the surroundings, or change the states of the
surroundings, conditional upon the internal states of the system.
\item Any macroscopic correlations between the current state of the
system and past states of its surroundings is screened off by the
existing correlations and interactions between the past time and the
present.
\end{enumerate}
Fortunately we do not need to construct explicit models for an
$IGUS$ or an $RIGUS$.  All we need to know is that either system
must be constructed out of the kind of operations described in the
previous sections.

It is now necessary to draw a distinction between the environmental
degrees of freedom of a heat bath, and the macroscopic states of the
surroundings that a computer might be correlated with. The set
$\{A_i\}$ refer to the internal logical states of the $IGUS$.  The
macroscopically distinct regions of the surroundings are $\{B_i\}$.
We represent the inaccessible regions of the environment by a
separate subsystem $\Omega_E$, which has no macroscopically
distinguishable subregions.  The overall state of the universe at
time $t$ is represented by $\Delta_t$.

\subsection{Growth in correlations}
Acquisition of knowledge is represented in the following terms. At a
time $t_1$ the computer is in the blank state represented by $A_0$,
while the surroundings are in one of the regions $B_i$.  The region
of state space is
\begin{equation}
\Delta_{i,t_1}=B_i \otimes A_0 \otimes E_{t_1}
\end{equation}
and the overall possible region is
\begin{equation}
\Theta_{t_1}=\cup_i B_i \otimes A_0 \otimes E_{t_1}
\end{equation}

 The acquisition of information
requires an evolution between $t_1$ and $t_2$ for which:
\begin{equation} \Delta_{i,t_2}=\phi^{(t_2)} \circ \phi^{-(t_1)}(\Delta_{i,t_1}) \subseteq B_i \otimes A_i
\otimes \Omega_E
\end{equation}
In an entropy increasing universe, we replace this by the coarse
graining $B_i \otimes A_i \otimes  \overline{E}_{i,t_2} \supseteq
\Delta_{i,t_2}$, for which
\begin{equation}
B_i \otimes A_i\otimes \overline{E}_{i,t_2} \subseteq B_i \otimes A_i \otimes
E^\prime_{i,t_2}
\end{equation}
for all $E^{\prime}_{i,t_2}$ that satisfy:
\begin{equation}
\Delta_{t_2} \subseteq B_i \otimes A_i \otimes E^\prime_{i,t_2}
\subseteq B_i \otimes A_i \otimes \Omega_E
\end{equation}
The overall region is
\begin{equation}
\Theta_{t_2}=\cup_i \phi^{(t_2)} \circ \phi^{-(t_1)}(B_i \otimes A_0
\otimes E_{t_1})
 \end{equation}
which has a coarse graining $\cup_i B_i \otimes A_i \otimes \overline{E}_{t_2}
\supseteq \Theta_{t_2}$, such that
\begin{equation}
\cup_i B_i \otimes A_i\otimes \overline{E}_{t_2} \subseteq B_i \otimes A_i
\otimes E^\prime_{t_2}
\end{equation}
for all $E^\prime_{t_2}$ that satisfy:
\begin{equation}
\Theta_{t_2} \subseteq \cup_i B_i \otimes A_i \otimes
E^\prime_{t_2} \subseteq \cup_i B_i \otimes A_i \otimes \Omega_E
\end{equation}

Now let us consider the reverse procedure, that would indicate the
existence of an $RIGUS$.  Start in $\Lambda_{i,t_1}=B_i
\otimes A_i \otimes E_{t_1}$ and perform the evolution
\begin{equation}
\Lambda_{i,t_2}=\phi^{(t_2)}_R \circ
\phi^{-(t_1)}_R(\Lambda_{i,t_1}) \subseteq B_i \otimes
A_0 \otimes \Omega_E
\end{equation}
This leads to the coarse graining
\begin{equation}
\Lambda_{i,t_2} \subseteq B_i \otimes A_0 \otimes
\overline{E^R}_{i,t_2}
\end{equation}
and the overall region
\begin{equation}
\Theta^{R}_{t_1}=\cup_i B_i \otimes A_i \otimes
E_{t_1}
\end{equation}
evolves into
\begin{equation}
\Theta^{R}_{t_2}=\cup_i \phi_R^{(t_2)} \circ \phi_R^{-(t_1)}(B_i
\otimes A_i \otimes E_{t_1})
 \end{equation}
which has a coarse graining $\cup_i B_i \otimes A_0 \otimes
\overline{E^{R}}_{t_2} \supseteq \Theta^{R}_{t_2}$, such that
\begin{equation}
\cup_i B_i \otimes A_0 \otimes \overline{E^{R}}_{t_2} \subseteq B_i
\otimes A_i \otimes E^{R\prime}_{t_2}
\end{equation}
for all $E^{R\prime}_{t_2}$ such that:
\begin{equation}
\Theta^{R}_{t_2} \subseteq \cup_i B_i \otimes A_i \otimes
E^{R\prime}_{t_2} \subseteq \cup_i B_i \otimes A_i \otimes
\Omega_E
\end{equation}

\subsubsection{Measures on marginals}
We now ask whether the requirement that an $RIGUS$ starts in a
correlated state, and removes those correlations, is less compatible
with an entropy increasing universe than an $IGUS$.  We will assume
that the internal
states of the $IGUS$ and $RIGUS$ have equivalent measures:
$\mu(A_0)=\mu(A_i)$.

First consider the measure of the initial states:
\begin{equation}
\mu(\Theta_{t_1})=\mu(\Theta^{R}_{t_1})
\end{equation}
An immediate consequence is that volume of state space arguments
will not be able to show preference for an $IGUS$ over an $RIGUS$ on
the basis of one or the other being simply more likely to occur at $t_1$.

From the measure preserving nature of the evolution of the $IGUS$ we
have
\begin{equation}
\mu(\Delta_{i,t_1})=\mu(\Delta_{i,t_2})
\end{equation}
while the coarse graining gives
\begin{equation}
\mu(B_i)\mu(A_0)\mu(E_{t_1}) \leq \mu(B_i)\mu(A_i)\mu(\overline{E}_{i,t_2})
\end{equation}
Similarly
\begin{equation}
\mu(\Theta_{t_1})=\mu(\Theta_{t_2})
\end{equation}
which when coarse grained gives
\begin{equation}
\sum_i \mu(B_i)\mu(A_0)\mu(E_{t_1}) \leq \sum_i
\mu(B_i)\mu(A_i)\mu(\overline{E}_{i,t_2}) \leq \sum_i
\mu(B_i)\mu(A_i)\mu(\overline{E}_{t_2})
\end{equation}

Using $\mu(A_0)=\mu(A_i)$, we get:
\begin{equation}
\mu(E_{t_1}) \leq \frac{\sum_i \mu(B_i) \mu(\overline{E}_{i,t_2}) }{\sum_i
\mu(B_i)}\leq \mu(\overline{E}_{t_2})
\end{equation}

While this might indicate an increase in entropy, we can easily get
similar results for the $RIGUS$.   The measures for the reverse
interaction are
\begin{equation}
\mu(\Lambda_{i,t_1})=\mu(\Lambda_{i,t_2})
\end{equation}
while the coarse graining gives
\begin{equation}
\mu(B_i)\mu(A_i)\mu(E_{t_1}) \leq
\mu(B_i)\mu(A_0)\mu(\overline{E^R}_{i,t_2})
\end{equation}
Similarly
\begin{equation}
\mu(\Theta^R_{t_1})=\mu(\Theta^R_{t_2})
\end{equation}
which when coarse grained gives
\begin{equation}
\sum_i \mu(B_i)\mu(A_i)\mu(E_{t_1}) \leq \sum_i
\mu(B_i)\mu(A_0)\mu(\overline{E^R}_{i,t_2}) \leq \sum_i
\mu(B_i)\mu(A_0)\mu(\overline{E^R}_{t_2})
\end{equation}
and $\mu(A_0)=\mu(A_i)$, gives:
\begin{equation}
\mu(E_{t_1}) \leq \frac{\sum_i
\mu(B_i)\mu(\overline{E^R}_{i,t_2}) }{\sum_i \mu(B_i)}\leq
\mu(\overline{E^R}_{t_2})
\end{equation}
It is clear that this $RIGUS$ interaction is just as entropy
increasing as the $IGUS$ interaction.  The direct growth in
macroscopic correlations of an $IGUS$ is no more indicative of
entropy increase than the reduction in macroscopic correlations
associated with an $RIGUS$.

\subsubsection{Micro- and macro-correlations}
The loss of microcorrelation with the environment is responsible for
the increase in entropy.  This happens both for the macroscopically
correlating interactions of an $IGUS$ and its reverse, $RIGUS$. What
of the macroscopic correlations themselves?  These are the
correlations which are supposed to be forbidden to develop within an
entropy decreasing universe.

While it is certainly true that the measure over the marginals
increases during information acquisition:
\begin{equation}
\sum_i \mu(B_i)\sum_j\mu(A_j) \geq \sum_i \mu(B_i)\mu(A_i)=\sum_i
\mu(B_i)\mu(A_0)
\end{equation}
(where we continue to assume $\mu(A_i)=\mu(A_0)$) this is a
qualitatively different kind of increase to that associated with
microcorrelations.  The coarse graining over the microcorrelations,
that corresponds to entropy increase, is associated with the
inaccessibility of these microcorrelations.  If the microscopic
correlations were still accessible (in the manner of a spin-echo
experiment) no entropy increase could be said to have occurred.

In the case of the macrocorrelations, however, it is essential that
the correlations be accessible.  It is precisely because the coarse
grained state is $\cup_i B_i \otimes A_i \otimes \overline{E}_{t_2}$ and
\textit{not} $\cup_i B_i \otimes \cup_j A_j \otimes \overline{E}_{t_2}$, that
the $IGUS$ is said to have information about its surroundings.  It
is the correlation that represents the information, that enables to
$IGUS$ to utilise that information in its interactions and future
behaviour.

The transition:
\begin{equation}
\cup_i B_i \otimes A_i \otimes E_{t_2} \rightarrow \cup_i B_i
\otimes \cup_j A_j \otimes E_{t_2}
\end{equation}
would represent a decorrelation, that would destroy the information
that the $IGUS$ held about the state of its surroundings.  So the
equivalent operation to the increase in entropy associated with
losing microcorrelations, is not associated with an acquisition of
information, but with its loss.

Let us consider the process by which such decorrelation occurs. In
an entropy increasing universe, each thermodynamically irreversible
operation increases the entropy of the surroundings and
environment.  Noise causes the switching of the computer's internal
states, or a switching (or change) of the environmental states.  An $IGUS$ must maintain
the relevance of its information by protecting against changes and
checking the accuracy of its information.  As the environmental
degrees of freedom become saturated, the existence of noise cannot
be protected against and decorrelation becomes irreversible. The
computer ceases to be able to function, as the universe approaches a
maximum entropy heat death.

Now it is precisely the fact that such irreversible decorrelation
does {\em not} occur (except on very large timescales), that
normally makes the information gathered useful.  The utilisation of
acquired information requires the existence of stable, accessible, macroscopic
correlations, so that the overall state, $\cup_i B_i \otimes A_i$ cannot be replaced by
the direct product of the marginal states, $\cup_i B_i \otimes \cup_j A_j$.  By contrast,
the increase in thermodynamic entropy is due to the loss of
microscopic correlations, which are presumed inaccessible.  This means that the
regions $\Delta_{i,t_2}=\phi^{(t_2)} \circ \phi^{-(t_1)}(B_i \otimes A_0 \otimes E_{t_1})\subseteq B_i \otimes A_i
\otimes \Omega_E$ can be replaced by the coarse grained, direct product of their marginal
states, $B_i \otimes A_i
\otimes \overline{E_{i,t_2}}$.  The role played by correlations in macroscopic
information and microscopic entropy turns out to be of a quite different nature.

\subsection{No interaction, no correlation}
There remains an intuition that, nevertheless, the kind of evolutions characterised as an $IGUS$ should still be more likely to occur that the kind of evolutions characterised by an $RIGUS$.  We will examine this further by considering a simple system, with two
states of the environment $B_i$ and two states of an $IGUS$, $A_i$.
If we suppose the system goes through the following stages:
\begin{equation}
A_0 \otimes B_0 \rightarrow A_0 \otimes \left(B_0 \oplus B_1\right)
\rightarrow \left(A_0 \otimes B_0 \right) \oplus \left(A_1 \otimes
B_1\right) \rightarrow \left(A_0 \oplus A_1\right)\otimes \left(B_0
\oplus B_1\right)
\end{equation}
Initially the system is in the low entropy, uncorrelated state.  The
environment evolves into one of two possible states.  The system
then measures the state of the environment, becoming correlated.
Eventually decorrelation leads to heat death.

The reverse, $RIGUS$, would involve:
\begin{equation}
A_0 \otimes B_0 \rightarrow \left(A_0 \otimes B_0 \right) \oplus
\left(A_1 \otimes B_1\right) \rightarrow A_0 \otimes \left(B_0
\oplus B_1\right) \rightarrow \left(A_0 \oplus A_1\right)\otimes
\left(B_0 \oplus B_1\right)
\end{equation}
At first sight, this evolution seems implausible.  We start with the
low entropy, uncorrelated state.  Correlations spontaneously appear.
The $RIGUS$ removes these correlations, before noise, once again,
leads to a heat death.

The problem in constructing a justification for eliminating the
$RIGUS$ evolution on entropic grounds is that:
\begin{equation}
\mu(A_0) \mu(B_0) \leq \mu(A_0)\mu(B_0) +\mu(A_1)\mu(B_1) = \mu(A_0)
(\mu(B_0) +\mu(B_1)) \leq (\mu(A_0) +\mu(A_1))(\mu(B_0) +\mu(B_1))
\end{equation}
the two intermediate states between the uncorrelated and the
decorrelated states can have the same measure.

Our intuition says that $A_0 \otimes \left(B_0 \oplus B_1\right)$
will occur first rather than $\left(A_0 \otimes B_0 \right) \oplus
\left(A_1 \otimes B_1\right)$.  The spontaneously correlated state
would require all initial states in $A_0 \otimes B_0$ to evolve into
either $A_0 \otimes B_0$ or $A_1 \otimes B_1$.  To achieve this it
is necessary for a correlated interaction to take place.  If it is
the case that at $t=0$, there is no correlation, and the two systems
do not interact (or share interaction with any combination of
intermediary systems) between $t=0$ and $t=\tau$, then
\begin{equation}\label{eq::maybefork}
\phi^{(\tau)}(A_0 \otimes B_0)=\phi^{(\tau)}(A_0) \otimes
\phi^{(\tau)}(B_0)
\end{equation}
Whatever else might be the case, such an evolution cannot possibly
induce a correlation.

If it seems surprising that such a conclusion can be drawn so
rapidly after the negative conclusion of the previous section, it is
important to notice the different.  The entropic argument was based
upon measures upon state space regions.  This argument is based upon
a restriction upon allowed evolutions of the combined system.

At first sight this might seem to provide the answer, neatly and
simply.  In an entropy increasing universe, the existence of
macroscopic correlation at some intermediate time requires the
existence of a macroscopic correlating interaction at an earlier
time.  By contrast, in an entropy decreasing universe, the existence
of macroscopic correlations at the intermediate time requires the
existence of a macroscopic decorrelating interaction in the future.
This appears to bear out Hawking's' claim that macrocorrelations must
decrease.

However, there are problems when one considers more complicated
situations than the two state systems considered here.  The no-interaction, no-correlation argument presented would rule out the development of microscopic correlations just as effectively as it rules out the development of macroscopic correlations.  In an
entropy increasing universe, microcorrelations \textit{must} develop, so it
seems that the restriction of equation \ref{eq::maybefork} is simply too
strong to represent the world. Once we allow microscopic correlations to be developing, it
is less clear what condition on the dynamics is necessary to ensure
an $RIGUS$ is less likely than an $IGUS$.

It might also seem implausible that the non-existence of an $RIGUS$
here and now, can genuinely be because of a boundary condition in the
remote past.  All the condition implies is that, given the existence
of a macroscopic correlation now, that there must have been, some
time between now and the start of the universe, a macroscopic
interaction.  It does not even guarantee that the systems which are
correlated, now, are the ones that interacted in the past - only
that there must have been an interaction in the past that has had
causal influences upon the two systems now.

In the time reversed situation, the future boundary condition is
supposed to prevent the operation of an $IGUS$.  However, all the
future boundary condition actually guarantees is that, at some point
in the future there must be a macroscopic interaction to remove the
correlation. It does not guarantee that this interaction must
involve the system currently correlated to its surroundings.  Given
the timescale involved for the future boundary condition to apply,
there seems a long way to go to show that a remote final condition
is sufficient to rule out the existence of $IGUS$ systems in an
entropy decreasing universe.  However, if true, this implies that a
remote boundary condition has a more direct effect upon possible
states now than just through the conditions given in Sections
\ref{ss:entinc} and \ref{ss:entdec}.

The argument now begins to resemble attempts to base the causal fork
asymmetry on entropic arguments\footnote{Note, this cannot have been
Hawking's intent, at least, as earlier in \cite{Hawking1994} he
speaks dismissively of causality and the arguments of
\cite{Reichenbach1971}.}. The literature on this topic is too large
to consider here (see
\cite{Reichenbach1971,Horwich1987,Alb01,Loewer2007} and
for criticisms see \cite{Pri96,Earman2006,Frisch2007}).  However, to
question how clear the argument is from a remote boundary condition
to situations now, we will simply consider two scenarios.  The first
will be Schulman's two time boundary condition, where both initial
and future boundary condition constraints exist. The second will be
a situation where a local entropy gradient exists, but without
either an initial or future boundary condition.  While these scenarios
may be regarded as implausible, their purpose is to examine if there
are gaps in the arguments based upon remote boundary conditions.

\subsubsection{Two time boundary conditions}
Suppose that we are in a two time boundary condition universe such
as Schulman proposes, but for which the thermalisation time is much
greater than half the lifespan of the universe.  In such a situation
one might find an overlap between the entropy increasing and
decreasing portions of space-time.  It may then be possible for a
complex system, operating in a thermodynamically reversible manner,
to operate in both temporal halves of the universe.  Now suppose
such a system is a computer is designed to work very close the
thermodynamic reversibility, and can swap from a power source
suitable for an entropy increasing universe to a power source
suitable for an entropy decreasing universe.

Why is it the case that, when the computer enters the entropy
decreasing timespan, it ceases to operate as an $IGUS$?  All we can
say is that, ultimately, any information it gathers, must be lost
again before the universe reaches its final low entropy state. That
seems to leave a large amount of time over which it is able to
function!  Of course, such a scenario also allows the possibility of
an equivalent $RIGUS$ existing in the entropy increasing period of
time.  The emergence of such a $RIGUS$ may be taken as an indicator
that a future boundary condition exists.  However, there seems no
direct reasoning, from thermodynamics, to tell us how far in the
future is such a boundary condition located.  If this is the case,
we equally cannot tell how long an $IGUS$ will be able to continue
to operate in an entropy decreasing universe.

\subsubsection{Asymmetry without boundary conditions}
The crossover, from a entropy increasing to decreasing universe,
raises additional problems, if we are to consider the interactions
between an $IGUS$ and an $RIGUS$ in the same region of space-time.
We can remove this problem by considering another, rather exotic,
situation, which questions whether a remote boundary condition could
possibly be responsible for the absence of $RIGUS$ systems.

Consider a system, identical to the solar system except in two
respects: the sun is not a sun, but a boundary that absorbs,
scatters and emits photons and particles into the solar system, with
exactly the same profile as our sun does; and around the solar
system (just around the Oort cloud) there is another closed
boundary, that absorbs, scatters and emits photons and particles
into the solar system with the same profile as the radiation
crossing a hypothetical surface enclosing our solar system.  Now
suppose that this completely enclosed system has been in this state
indefinitely far into the past, and will be in this state
indefinitely far into the future.

Such a system is explicitly time asymmetric.  The profile of the
radiation being absorbed, scattered and emitted on the two
boundaries is quite different when viewed in a time reverse
direction.  In a normal time direction the solar boundary emits low
entropy radiation, some of which falls upon an earth-like planet and
is reradiated in a higher entropy form.  Most of the solar
radiation, along with most of the earthly re-radiation is eventually
absorbed by the Oort boundary, which radiates a negligibly small
amount of radiation back (largely concentrated at small points)
apart from a roughly symmetric emission and absorbtion of radiation
at the cosmic microwave background frequency.  Reversing the time
direction will produce a quite different profile of emission and
absorbtion on the two boundaries.

Let us ignore issues, such as the question of the long term
stability of the solar system and so forth, which are not directly
relevant to the present day thermodynamics of our solar system. For
much of the history of life on our earth, there has been a
reasonably stable non-equilibrium state, maintained by the local
entropy gradient between the radiation falling on earth, from our
sun, and re-radiated out again.  The enclosed solar system will be
in a stable non-equilibrium state much like our solar system,
including the earth-like planet.  It would seem reasonable to expect
conditions on the earth-like planet to resemble conditions on our
earth.

The principal argument of this paper has been that there appears
nothing in the thermodynamics of the local conditions on the
earth-like planet that prevents the existence of an $RIGUS$.  The
no-interaction no-correlation argument suggests that a remote
initial boundary condition prevents it on our earth.  However, in
the enclosed solar system, there is no remote initial boundary
condition.  If an $RIGUS$ is still not possible in the enclosed
solar system, it must be the case that there is something about the
local entropy gradient that prevents it, rather than an initial
boundary condition.

If the remote initial boundary condition has an influence on the
state of our earth only through the entropy difference between the
incoming and outgoing radiation, then the cause of the absence of an
$RIGUS$ on earth must be the same as on the enclosed earth-like
planet.  We have found no explanation in terms of the local entropy
gradient to prevent an $RIGUS$, so if the local entropy gradient
screens off the effect of a remote initial boundary condition, then
such a condition cannot provide an argument against the existence of
an $RIGUS$ on our earth.

Alternatively, the remote initial boundary condition may have a
direct effect on the conditions on earth today that is not screened
off by the local entropy gradient.  In this case it may prevent the
existence of an $RIGUS$ on our earth, but leaves the possibility of
an $RIGUS$ on the enclosed earth-like planet.  It is hard to see
what kind of process could supply such a direct effect, or how this
would lead to conditions being so radically different on the
enclosed earth-like planet, but one possibility might be the
asymmetry of electromagnetic radiation, between advanced and
retarded waves.

\section{NESS, not QSES. Complexity, not information}
Any process that is a sequence of Quasi-Static Equilibrium States
(QSES) can, in principle, be connected by thermodynamically
reversible processes (it is this that enables us to determine the
entropy difference between them). Let us consider a specific
example: the paradigmatic ice cube melting in a glass of water, and
the film of this being run backwards.

There is nothing about the two states: an ice cube in glass; and a
glass of water; that tells us one must come before the other.  It is
entirely possible in an entropy increasing universe, for the ice
cube to be in the future of the glass of water.  There are entropy
increasing processes by which a glass of water can be turned into a
glass containing an ice cube.  In the limiting case, of reversible
quasistatic processes, we can go back and forward between ice cube
and water, thermodynamically reversibly.

The same is equally true in an entropy decreasing universe.  In such
a universe there would also be entropy decreasing processes by which
glasses of water could be converted into ice cubes in glasses and
ice cubes in glasses converted into glasses of water.

The asymmetry in the process, with which we are familiar, is not the
fact that an ice cube is succeeded by water, but is in the process
by which it happens.  It is the non-equilibrium nature of the
process that reveals the entropic direction.  It is the fact that
the ice cube is in the process of {\em melting} that tells us the
`correct' direction of the film.

The generalisation of the arguments Section \ref{ss:trevseq} is
that any process, which can be defined solely in terms of a
(deterministic or probabilistic) succession of QSES, can occur in an
entropy increasing universe and in an entropy decreasing universe.
What distinguishes the two universes is not a possible succession of
QSES, but rather the processes by which the transitions between the
states can take place.  This suggests that, if one is to find
connections to an entropic arrow of time, we should not be looking
at the QSES that are the thermodynamically reversible limit for information
processing systems. Any process which can be defined solely in terms of such
states can occur in either entropic direction.

The existence of Non-Equilibrium Steady States (NESS), on the other
hand, are not time symmetric.  Complex biochemical structures that
arise in far from equilibrium conditions are associated with
fundamentally time asymmetric, entropy increasing processes. The
time reverse of these processes in entropy decreasing universes will
lead to a different sequence of NESS, entropy decreasing processes.
These complex structures are also the building blocks from which the
biological processes are constructed that are necessary to house the
information gathering and utilising systems.

A generalisation of this may be conjectured: any time asymmetry that
is supposed to be a consequence of the thermodynamic time asymmetry,
cannot be expressed solely in terms of sequences of QSES.  If we are
to find stable states whose time asymmetry is a consequence of
thermodynamics, their properties must come from NESS, not QSES. This
suggests that the ideas of complexity, rather than information, are
needed.

\section{Conclusion}
The argument of this paper is that an arrow of time associated with
information processing systems cannot be deduced from thermodynamic
arguments.  The thermodynamic arrow is insufficient to entail the
computational arrow.  Any sequence of logical operations in an
entropy increasing universe is physically possible in an entropy
decreasing universe.  Landauer's Principle, as it is commonly
stated, assumes statistical mechanical principles that are
equivalent to being in an entropy increasing universe.  If one
changes those assumptions, so that one is in an entropy decreasing
universe, a critical inequality in Landauer's Principle in reversed.
The physical implementation of logical operations, which increase
entropy, do so, not by virtue of any inherent properties of the
logical operation, but by virtue of being in an entropy increasing
universe. If the same logical operation is performed in an entropy
decreasing universe, it is entropy decreasing.  As a result, entropy
decreasing universes are not inherently hostile to the acquisition,
persistence or utilisation of information.

In principle, the operation of acquiring information can be made
thermodynamically reversible.
 This is precisely one of the main
insights of Landauer's work on the thermodynamics of computation: a
measurement can take place without generating heat (see
\cite{LR90,LR03} and many references within).
\begin{quote}
Landauer's principle, while perhaps obvious in retrospect, makes it
clear that information processing and acquisition have no intrinsic,
irreducible thermodynamic cost\cite{Ben03}
\end{quote}
If the acquisition of information can take place in a
thermodynamically neutral manner, it can take place in an entropy
decreasing as easily as an entropy increasing universe.

While any information gathering and utilising system will ultimately
cease to function in an entropy decreasing universe that reaches a
final extremal entropy state, this doesn't seem sufficient to rule
out such systems\footnote{Quite aside from the fact that it must
also ultimately cease to function in an entropy {\em increasing}
universe.}. Firstly, the decrease in entropy is due to the
decorrelation that comes about from losing microcorrelations. It is
of a different kind to the macrocorrelations that arise during the
acquisition of information.  Secondly, on the timescales during
which information gathering and utilising systems work, between the
low or high entropy extremal starting and ending points, there seems
nothing to directly prefer $IGUS$ over $RIGUS$.  Thirdly, if the
effect of the initial or future boundary conditions is screened off
by the local entropy gradient, the no correlation, no interaction
argument does not seem to be applicable, as such an entropy gradient
can exist in a situation with no initial or final boundary
condition.

The suggestion is made that an entropic arrow of time will never be
found in processes that can be defined solely in terms of a
succession of Quasi-Static Equilibrium States.  Information
processing can be so defined.  If the psychological arrow of time is
to be aligned with the thermodynamic arrow, it cannot be through the
information processing properties of the brain.  It may be through
the biochemical structures that arise in Non-Equilibrium Steady
State processes, but if so, it is certainly not through any
information processing characterisation of such structures. This
would seem to imply that at least one aspect of conscious experience
cannot be logically supervenient on the states of a computer.  If
instead the psychological arrow of time does indeed arise out of
information processing properties, this would mean that the
psychological arrow is logically independent of the thermodynamic
arrow of time.

\begin{acknowledgements}
I would like to thank Avshalom Elitzur,
Steve Weinstein and Jos Uffink for interesting discussions.

Research at Perimeter Institute for Theoretical Physics is supported
in part by the Government of Canada through NSERC and by the
Province of Ontario through MRI.
\end{acknowledgements}

\bibliographystyle{spbasic}      



\end{document}